\documentclass[5p,preprint]{elsarticle}
\graphicspath{{}} 
\usepackage{graphicx}

\usepackage{subfigure}
\usepackage{caption}

\usepackage{url}

\def\ptstar{\ensuremath{\mathrm{p_T^*}~}}

\def\mum{\ensuremath{\mu\rm m}}

\def\beq{\begin{equation}}
\def\eeq{\end{equation}}
\def\beqa{\begin{eqnarray}}
\def\eeqa{\end{eqnarray}}

\def\xipil{\ensuremath{\Xi^- \to \pi^- \Lambda~}}

\def\xim{\ensuremath{\Xi^-~}}
\def\om{\ensuremath{\Omega^-~}}

\title{Considerations on \xim reconstruction in LHCb.}
\author[rvt]{ F.M.~Brochu\corref{cor1}}
\ead{brochu@hep.phy.cam.ac.uk}
\cortext[cor1]{Corresponding author.}
\address[rvt]{HEP group, Cavendish Laboratory,JJ Thomson Avenue,\\ Cambridge CB3 0HE, United Kingdom }

\begin{document}

%
%
\begin{abstract}
This paper describes an alternative method of charged hyperon reconstruction applicable to the LHCb experiment. It extends the seminal work of the FOCUS collaboration \cite{FOCUS} 
to the specific detector layout of LHCb and addresses the reconstruction ambiguities reported in their earlier work, leading to improvements in the reconstruction efficiency for the specific cases of \xim and \om baryon decays to a charged meson and a $\Lambda$ baryon. 
\end{abstract}
\begin{keyword}
hyperon reconstruction \sep LHCb \sep kink 
\end{keyword}
\maketitle
\section{Introduction}
The LHCb experiment  has a unique opportunity to study heavy baryons produced abundantly in proton-\linebreak proton collisions at the LHC, as it is able to efficiently trigger \cite{Trigger} on both muonic and purely hadronic final states and to reduce the high level of hadronic background thanks to its particle identification system \cite{RICH}. Although it has not failed to provide regular publications on b baryon spectroscopy, only two papers so far have been devoted to final states containing a \xim or an \om \cite{ObXibLife,LbObXibMass}.\\
It was mentioned in \cite{LHCbOb} that the reconstruction of charged hyperons, and in particular the \xim and \om baryons, was particularly inefficient in the LHCb detector because of long-lived, secondary lambda baryons decaying outside its acceptance.\\
This paper presents an alternative reconstruction method aiming to address this inefficiency and to promote the study of decay channels containing a \xim or a \om at LHCb.  We start with the experiment itself and the way it reconstructs the decay channels mentioned above, then we describe the method we aim to apply and finish by a section about the potential of the method and a discussion about its advantages and disadvantages. 

\section{Reconstruction of \xim in LHCb.}
\begin{figure}
\includegraphics[width=9cm,height=6cm]{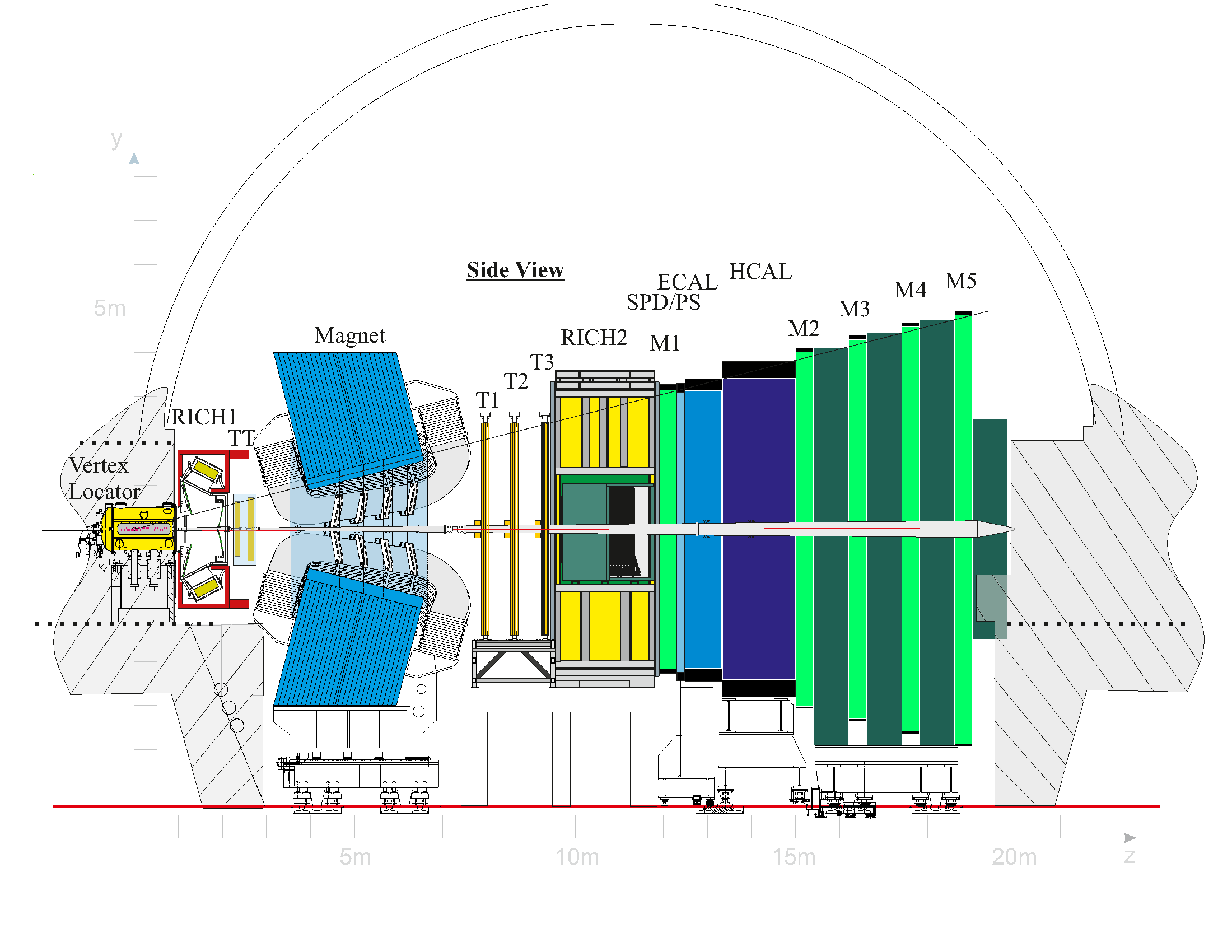}
\caption{LHCb detector layout. The interaction point is on the left, inside the VELO detector. Other tracking volumes of interest to this study are, from the left to the right, the upstream tracker (TT) and  downstream tracker (T1,T2 and T3 or "T stations").}
\label{fig:LHCb}
\end{figure}

\begin{figure}
\includegraphics[width=9cm,height=5cm]{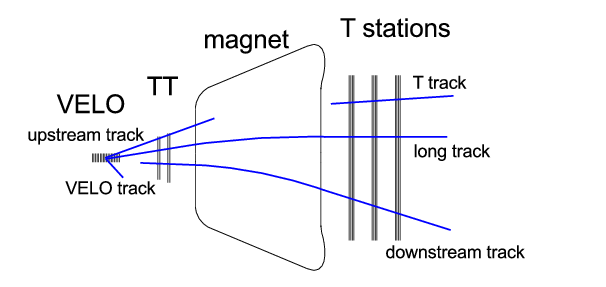}
\caption{Schematic representation of LHCb track types. This study uses "Long" and "Downstream" tracks, as well as segment seeds called "VELO tracks" and "T tracks".}
\label{fig:tracking}
\end{figure}

The LHCb detector \cite{LHCb1,LHCbperformance} is a single-arm forward spectrometer covering the pseudo-rapidity range $2 < \eta < 5$ designed for the study of heavy particles containing b or c quarks. Its layout and different components are shown in Figure \ref{fig:LHCb}.\\
The detector includes a high-precision tracking system consisting of a silicon-strip vertex detector surrounding the pp interaction region (labelled in Figure 1 as "VELO"), a large-area silicon-strip detector (TT) located upstream of a dipole magnet with a bending power of about 4Tm, and three stations of silicon-strip detectors and straw drift tubes placed downstream of the magnet ("T stations"). The particle identification system mentioned above is for-med by two ring-imaging Cherenkov detectors (RICH1 and RICH2).\\
Charged particles trajectories are reconstructed through the different elements of the tracking system. Standalone seed tracklets are first formed from hits in each element, before being combined in a global track fit. It is important to note that no tracking element is located inside the dipole magnet, and therefore accurate momentum reconstruction will require the combination of segment seeds from both side of the magnet. Fringe fields outside the magnet allow the reconstruction of the momentum of the "T tracks", but this will be neglected in this analysis due to their reduced momentum resolution \cite{Schiller}. With only two tracking sub-systems upstream of the magnet, there are only two types of track used in LHCb analyses: the "Long" tracks, with hits from all tracking subsystems, and the "Downstream" tracks, with no hit from the VELO. Tracks and segment seeds are represented in Figure \ref{fig:tracking}.
The \xim is a long-lived, charged hyperon that decays into a charged pion and a $\Lambda$. The $\Lambda$ itself is long-lived and is reconstructed through its decay into a proton and a charged pion of opposite charge. The reconstruction of the charged hyperon can only proceed if all three charged particles have their momentum measured, which means that they need to form either "Long" or "Downstream" tracks. In \cite{Marki}, the analysis recognises three types of combinations leading to the reconstruction of a \om or \xim: "LLL" where all three particles leave "Long" tracks, "LDD", where the proton and pion from the $\Lambda$ decay are produced outside the VELO and are reconstructed as "Downstream" tracks, and "DDD" where all three particles are reconstructed as "Downstream" tracks.  We hereby refer to this reconstruction method as "3-tracks" sum.\\
Needless to say, as the $\Lambda$ is produced downstream of the charged hyperon and decays even further from the interaction point, there will be cases where the $\Lambda$ decay products will be created after the TT, the last tracking station before the magnet, and it will not be possible to reconstruct their momentum.   
\section{Reconstruction of \xim as kink.}
\begin{figure}
\includegraphics[width=9cm,height=5cm]{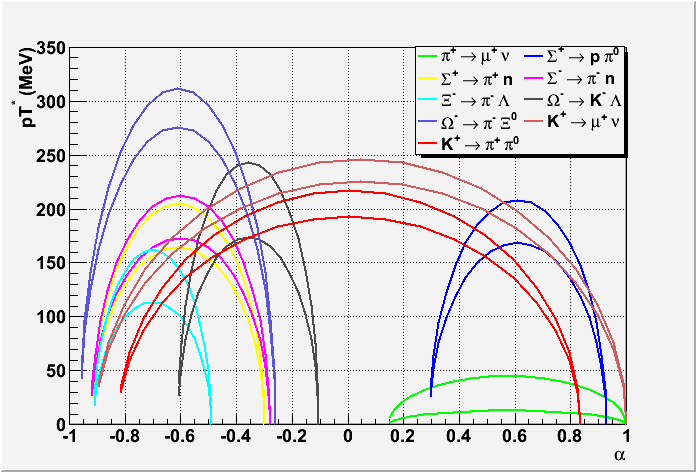}
\caption{Armenteros-Podolanski diagram for "kinks". The lower and upper boundaries of each band are 20 MeV below and above the central mass to account for potential detector resolution effects.  }
\label{fig:kinksAP}
\end{figure}

 The reconstruction method we propose here is not new: kink reconstruction was tried and documented by the FOCUS collaboration in \cite{FOCUS}. 
Long-lived ($c\tau > 1$cm), charged particles leave a measurable primary track before decaying in apparatus like FOCUS or LHCb thanks to the forward geometry of the detector, optimised for the reconstruction of highly boosted particles. As the charged particle decays to another charged particle and at least one neutral particle, the secondary charged particle forms another measurable track in the detector with a different direction, leading to a kink.\\
The FOCUS collaboration has identified 8 different decay channels corresponding to simple kink topologies, which we represent in the Armenteros-Podolanski diagram in Figure \ref{fig:kinksAP}.

The Armenteros-Podolanski variables used in Figure 3 are defined as follows:
\begin{itemize}
\item on the x axis: $\alpha$, the asymmetry of the longitudinal projection of the daughter momenta with respect to the line of flight of the mother particle.\\
$\alpha$ = $\frac{p_L^{ch} - p_L^0}{p_L^{ch} + p_L^0}=\frac{2 p_L^{ch}}{P}-1$, where the superscripts $ch$ and $0$ are used for the charged and neutral daughter respectively, and the capital $P$ used for the mother particle momentum.
\item on the y axis: \ptstar, the transverse projection of the charged daughter momentum with respect to the line of flight of the mother particle.
\end{itemize}
 
One can immediately see from this busy diagram the importance of reconstructing the momentum magnitude of the mother particle. Without it, we cannot reconstruct $\alpha$, and therefore we cannot tell from the kink topology alone which decay channel is being observed.
Unfortunately, neither FOCUS nor LHCb have the possibility of reconstructing the momentum of the mother particle before it decays, as the IP region is not covered by any substantial magnetic field allowing such measurement, so one has to use the daughters' data to distinguish between kinks. For both \xim and \om baryons, the neutral daughter is a $\Lambda$ baryon, which leaves a distinctive topological signature of two short, intersecting tracks (or "V0") in the tracking detectors, so all we need to distinguish between \om and \xim decays to $\Lambda h^-$ are the ring-imaging Cherenkov detectors that will tell us if $h^-$ is a pion or a kaon. But unfortunately, this is not the end of the story as the $\Lambda$ can decay beyond the TT chambers, leaving no "Downstream" track to measure its momentum.

\section{Application to LHCb.}
\begin{figure}
\includegraphics[width=9cm,height=5cm]{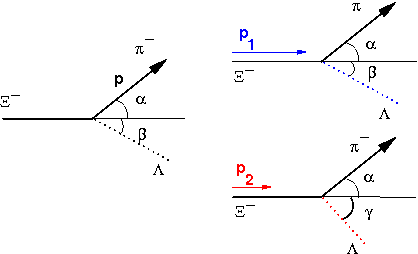}
\caption{Left: illustration of the complete reconstruction of the \xim kinematics with the full momentum of the charged daughter (arrow) and the flight directions of the \xim (plain line on the left) and the $\Lambda$ (dashed) . Right: The two possible solutions given the mother direction and charged daughter full momentum from the right-hand side decay (kinked, plain arrow). Top: high \xim momentum $p_1$ (narrow angle $\beta$). Bottom: low \xim momentum $p_2$ (wider angle $\gamma$). Angles are magnified for the sake of clarity.}
\label{fig:kinks}
\end{figure}

We have seen in the previous section that, even by reconstructing the \xim as a kink, we still need to consider the $\Lambda$ decay in order to remove the two-fold ambiguity arising from the unmeasured mother track momentum. However, we do not need to reconstruct the $\Lambda$ momentum: with the mother particle flight direction known and materialised as a VELO track, and the charged daughter full momentum vector (direction and magnitude) reconstructed from the matching "Long" or "Downstream" track, all we need to know is the direction of flight of the Lambda, as demonstrated in Figure \ref{fig:kinks}. 
From Figure \ref{fig:kinks}. with the momentum of the daughter track fully reconstructed (direction and magnitude), the momentum magnitude of the $\Lambda$ is reconstructed as $p_\Lambda=p \frac{sin(\alpha)}{sin(\beta)}$ from transverse momentum conservation. With the momentum direction of the $\Lambda$ known, a simple vector sum returns the momentum of the mother.

This direction is given by the line joining the production and decay points of the $\Lambda$, with the production point reconstructed as kink vertex. The decay point of the $\Lambda$ will be reconstructed as the intersection of two "T tracks" to maximise the reconstruction efficiency. \\
Other kinematic properties of the decay should be applied to reduce background contributions:
\begin{itemize}
\item momentum conservation: the decay is a two-body decay, so the vertex must sit cleanly in the kink plane, defined by the flight direction of the mother and charged daughter at the kink vertex.
\item kink mass. As mentioned by FOCUS, there is only a two-fold ambiguity to the mother momentum magnitude from the kink reconstruction. These two solutions translate into two possible directions for the $\Lambda$, with two different emission angles with respect to the flight direction of its mother as shown schematically in Figure \ref{fig:kinks}.  
\end{itemize}
The second point is quite important, as we need to be able to distinguish clearly between the two solutions to perform a correct momentum assignment to the charged hyperon. We found the average angular difference between these solutions for a \xim with a boost of 35 to be 2 mrad.\\
We evaluated both reconstruction methods' efficiencies on a Monte-Carlo simulation of the decay $\Xi_b^- \to J/\Psi \Xi^-$,$\xipil$ that was studied by LHCb in \cite{ObXibLife,LbObXibMass,Marki}, using the PYTHIA generator \cite{PYTHIA} to simulate proton-proton collisions with a centre of mass energy of 14 TeV. Generated $\Xi_b^-$ were forced to decay in the channel of interest using the software package EvtGen \cite{EVTGEN}. We used the published LHCb acceptance $2<\eta<5$ and a lower momentum cut of 2 GeV for all particles involved in the decay to define our sample of "decays within acceptance". \\
We used the geometrical descriptions of the upgraded LHCb tracking detectors defined in \cite{Upgrade1,Upgrade2} to define our subsets of "reconstructible" tracks and reconstruction classes below. The differences with the detector layout shown in Figure \ref{fig:LHCb} is described in \cite{Upgrade1,Upgrade2}. The main differences relevant to this study are an improved resolution for the upgraded VELO and SciFi detector, the latter replacing the "T stations". 
The reconstruction efficiencies are the fraction of decays within acceptance that are also "reconstructible" with one of the two methods described above, the "3-tracks sum" used by LHCb in  \cite{ObXibLife,LbObXibMass,Marki} and the "kink" method we proposed.\\
Using the upgraded detector layout and the Monte-Carlo simulation of the decay $\Xi_b^- \to J/\Psi \Xi^-, \xipil$ , we define several, non-overlapping, reconstruction classes.
\begin{itemize}
\item "Short-lived \xim" ($\sim 30$ \% all decays). The \xim decays before reaching the third VELO station needed to leave a "reconstructible" VELO track \cite{Upgrade1}. Only the traditional "3-tracks" method can be used here.
 \item "Medium range" ($\sim 35$ \% all decays). The \xim leaves a reconstructible track in the VELO. Its charged daughter is reconstructed as either a "Long" or "Downstream" track. Its neutral daughter decays before the TT detector, allowing for these decays to be reconstructed indifferently as either a "kink" or a "3-tracks" resonance.
\item "Long range" ($\sim 15$ \% all decays). The \xim leaves a reconstructible track in the VELO, but the daugther $\Lambda$ decays beyond the first TT layer, so its own decay products are not reconstructible as "Downstream" tracks, only as "T tracks". These decays can only be reconstructed as kinks.
\item "Out of acceptance" ($\sim 20$ \% all decays). These decays cannot be reconstructed in any way, as either the \xim decays past the first TT layer, and the momentum scale cannot be measured from the charged daughter of the \xim, or the $\Lambda$ escapes detection.
\end{itemize}
The "Long range" category is only useful if we are able to resolve the "two-fold" ambiguity attached to the "kink" reconstruction. To do so, we proposed to look for the $\Lambda$ decay vertex between the TT and the SciFi detectors by looking for two intersecting SciFi segments. As the $\Lambda$ may decay in a region with significant magnetic field affecting the trajectories of its charged daughters, we conservatively considered 2D vertexing in the non-bending plane.
Using the Monte-Carlo truth data, we compared the effective separation of the two possible decay vertices for the $\Lambda$ with the vertex resolution achieved by using two intersecting, straight SciFi segments to reconstruct the $\Lambda$ decay vertex. \\
Doing this in a decay by decay basis, we found that we are unable to resolve only $\sim$ 10 \% of the decays of the "Long range" category (or 1.5 \% of all decays). This is due to the superior resolution of the SciFi tracker (100 $\mum$ single-hit, transverse resolution). The "kink" method is therefore a viable one to recover \xim decays where the $\Lambda$ decays beyond the TT.
\section{Discussion}
We can see from here that the two reconstruction methods, "kink" and "3-tracks" sum offer significant overlap ($\sim 35$ \% of all decays in acceptance) and that the comparison of each method on their own is not favourable to the "kink" method. So we suggest to use the two methods together, as the benefits from the "kink" method does not stop at the "Long range" decays.\\
In the "Medium range" class, the benefits of reconstructing the \xim track along with the 3 tracks of its decay products are the following:
\begin{itemize}
\item The \xim decay vertex (or "kink" vertex) is constrained with a high resolution track. This allows for potential improvements in the \xim mass and momentum resolutions.
Although it is beyond the scope of this study to evaluate accurately how much resolution can be gained, we can get a rough idea by smearing the Monte-Carlo values of the \xim decay used in the kink reconstruction. Smearing was done using detector resolution quantities listed in \cite{Upgrade1,Upgrade2}.
We found a \xim mass resolution of $\sim$ 15 MeV, dominated by the uncertainty on the \xim track direction. This is due to the fact that the reconstructed \xim track is predominantly a short track ($\sim 100-200$ mm) with a minimum number of hits.
\item The reduction of combinatoric "V0 + secondary\linebreak track" background, by showing that there is a detector reading where the \xim flies by.
\item The \xim {\bf production} vertex is also better constrained. In this particular case, the \xim track is added to the $J/\Psi$ decay products ($J/\Psi \to \mu^+ \mu^-$) to form a three track vertex instead of a two track one. 
\end{itemize}

\section{Conclusions}
 We have presented a complementary method to \xim  \linebreak reconstruction from its decay products alone, using the track left by the boosted \xim in the LHCb Vertex Locator to form a kink or over-constrain the reconstruction. The kink reconstruction method extends the works of the FOCUS collaboration by providing a way to resolve the two-fold ambiguity associated to the kink reconstruction by looking for the $\Lambda$ decay vertex in LHCb's downstream tracker without having to reconstruct the $\Lambda$ momentum explicitly. We have also shown that its replacement in the coming upgrade, the SciFi tracker, will have the resolution needed to perform this task.\\
We hope that this study will encourage many LHCb physicists to take on the study of decay channels containing a \xim or a \om, as the results of this study apply in principle to \om decays to $\Lambda K^-$ as well (although with different efficiencies, coming from kinematics and lifetime differences).
\section*{Acknowledgments}
The author would like to thank Dr Susan Haines, Pr Val Gibson and Dr Chris Jones for reading and correcting the manuscript. 

\end{document}